%Paper: hep-th/9307160
%From: MLEBLANC@LPS.UMONTREAL.CA
%Date: Mon, 26 Jul 1993 15:14:11 -0400 (EDT)

\magnification=1200
\rm
\baselineskip = 12 pt plus 1 pt  minus 1 pt
\tolerance=1000
\pretolerance = 5000
\voffset = 2.4 true cm
%\hoffset = 3 true cm
\vsize = 22 true cm
\hsize = 17 true cm

\def\ick{\eqalignno}

\def\overrightarrow#1{\vbox{\ialign{##\crcr
    \rightarrowfill\crcr\noalign{\kern-1pt\nointerlineskip}
    $\hfil\displaystyle{#1}\hfil$\crcr}}}
\def\overleftarrow#1{\vbox{\ialign{##\crcr
    \leftarrowfill\crcr\noalign{\kern-1pt\nointerlineskip}
    $\hfil\displaystyle{#1}\hfil$\crcr}}}

\def\ick{\eqalignno}
\def\sqr#1#2{{\vcenter{\hrule height.#2pt
   \hbox{\vrule width.#2pt height#1pt \kern#1pt
       \vrule width.#2pt}
         \hrule height.#2pt}}}

\def\_{^{}_}

\topinsert

\endinsert

\font\small=cmr9
\def\tepsilon{\tilde\epsilon}
\def\bof#1{\hbox{}}
%\headline{\hfill {\bf PRELIMINARY}}

\centerline {\bf Two Dimensional Gauge Theoretic Supergravities$^\star$}

\vskip 12pt

\centerline {{\bf D. Cangemi$^{(1)}$ and M. Leblanc$^{(2)}$ }}

\vskip 12pt

{\baselineskip = 12pt plus 1pt minus 1pt
\it
\centerline { {\rm $(1)$} Center for Theoretical Physics}
\centerline { Massachusetts Institute of Technology}
\centerline { Cambridge, MA 02139}
\centerline {U.S.A.} }

\vskip 10pt
{\baselineskip = 12pt plus 1pt minus 1pt
\it
\centerline { {\rm $(2)$} Laboratoire de Physique Nucl\'eaire and}
\centerline {Centre de Recherches Math\'ematiques }
\centerline { Universit\'e de Montr\'eal}
\centerline { C.P. 6128, succ. A}
\centerline { Montr\'eal, (Qc), Canada}
\centerline { H3C 3J7} }
\vskip 10pt
\centerline {\bf Submitted to: {\it Nuclear Physics B}}
\vskip 10pt
\centerline {\bf Abstract}
\vskip 15pt

We investigate two dimensional supergravity theories, which can be built from
a topological and gauge invariant action defined on an ordinary surface.
We concentrate on four models. The
first model is the $N=1$ supersymmetric extension of Jackiw-Teiltelboim
model presented by Chamseddine in a superspace formalism. We complement the
proof of
Montano, Aoaki, and Sonnenschein that this extension
is topological and gauge invariant, based on the graded
de Sitter algebra. Not only do the equation of motions correspond to the
supergravity ones and gauge transformations encompass local
supersymmetries, but also we identify the $\int \langle \eta, F\rangle $-theory
with the superfield formalism action written by Chamseddine. Next, we show
that the $N=1$ supersymmetric extension of string inspired two dimensional
dilaton gravity put forward by Park and Strominger is a theory that satisfies a
non-vanishing curvature condition and cannot be written as a $\int\langle
\eta,F\rangle$-theory. As an alternative, we propose two examples of
topological and
gauge invariant theories that are based on graded extension of the extended
Poincar\'e algebra and satisfy a vanishing curvature condition. Both
models are interpreted as supersymmetric extensions of the string inspired
dilaton gravity.
\vfill
\line{CTP\#2224}
\line{UdeM-LPN-TH-93-167}
\line{CRM\#1897 \hfill July 1993}
\vskip 10pt
\hrule
\vskip 5pt
{\small $^\star$This work is supported in part by funds provided by the U.S.
Department of Energy under contract \#DE-AC02-76ER03069, by the Fonds du
450e de l'Universit\'e de Lausanne (D.C.), by the Natural Sciences and
Engineering Council of Canada, and by the Fonds pour la Formation
de Chercheurs et l'aide \`a la Recherche (M.L.).}
\eject
\medskip
\centerline {\bf R\'esum\'e }
\medskip
Nous \'etudions dans cet article les
th\'eories de la supergravit\'e \`a (1+1)-dimensions
qui s'\'ecrivent comme une th\'eorie topologique et invariante de
jauge. Nous nous concentrons sur quatre mod\`eles.
Le premier mod\`ele est l'extension
supersym\'etrique $N=1$ du mod\`ele
de Jackiw-Teitelboim d\'evelopp\'ee par Chamseddine dans un formalisme de
super-espace. Nous compl\`etons la preuve de Montano, Aoaki, et Sonnenschein
qui en montre le caract\`ere topologique et invariant de jauge, bas\'e
sur l'alg\`ebre de de Sitter : Non seulement les \'equations du
mouvements correspondent \`a celles de la supergravit\'e usuelle et
les transformations de jauges sont celles de la supersym\'etrie locale,
mais en plus l'action de Chamseddine est identifi\'ee \`a une action
$\int \langle\eta, F\rangle$.
Nous prouvons \'egalement que l'extension
supersym\'etrique $N=1$ du mod\`ele
de gravit\'e dilatonique, inspir\'e par la th\'eorie
des cordes,
propos\'ee par Park et Strominger est une th\'eorie qui satisfait une
condition de courbure non nulle et ne peut pas s'\'ecrire comme une th\'eorie
$\int \langle\eta, F\rangle$. On propose ensuite deux exemples
de th\'eories topologiques et invariantes de jauge bas\'ees sur des
g\'en\'eralisations de l'alg\`ebre de Poincar\'e \'etendue et qui satisfont
une condition de courbure nulle. Les deux mod\`eles sont interpr\'et\'es
comme des extensions supersym\'etriques de th\'eories de gravit\'e
dilatonique inspir\'ee par les th\'eories bidimensionelles de corde.
\vfill
\eject
\headline{\hfill}
\noindent {\bf Introduction}
\medskip
Recently, there has been much interest for two dimensional gravity
theories,$^{1,2,3}$
such as the Jackiw-Teitelboim (JT) model$^1$ and the string inspired dilaton
gravity (SI) model.$^4$
The reasons are numerous and among them, it is the presence of
black holes that is attractive.$^{4,5}$
Their resemblance to Einstein gravity theory
does not end with the presence of black holes. Gravitational collapse,
propagation of gravitational waves and Newtonian expansion are also
present.$^6$
Two dimensional gravity theories are therefore an interesting arena to
explore features of gravity without the difficulties encountered in the four
dimensional world.

Both of the above theories (JT/SI) share an interesting property. They both can
be written as topological gauge field theories.$^{7,8,9}$ It is
interesting because
gravity in four dimensions has not been successfully written as a gauge theory.
In two dimensions, it provides another way to analyse gravity theories. This
was fruitful for the three dimensional case: general
relativity theory in (2+1) dimensions was shown to be equivalent to a
Chern-Simons gauge theory based on the Poincar\'e algebra$^{10}$
and the Ba\~nados,
Teitelboim, and Zanelli black hole solution in (2+1) dimensional anti-de Sitter
spacetime$^{11}$ fits in a gauge formulation.$^{12}$ A similar study of
Chern-Simons supergravity as a supersymmetric gauge theory is presented in
Ref.~13.

The first model of 2d-gravity was proposed some time ago by Jackiw and
Teitelboim.$^1$ It is obtained by dimensionally reducing the usual
Einstein-Hilbert action in (2+1) dimensions,
$$
S_{\rm JT}= {1\over 4\pi}
\int d^2x {\sqrt {-g}}\;\; \eta \Bigl ( R - \lambda\Bigr )
\eqno (1.1)
$$
the scalar curvature $R$ is equated to a
cosmological constant $\lambda$ through a Lagrange multiplier $\eta$.

The second model appeared recently and is inspired by string theory (SI)
when restricted to a (1+1) dimensional target space$^4$
$$
{\overline S}_{\rm SI}={1\over 4\pi}
\int d^2x\; {\sqrt {-{\overline g}}} \;e^{-2\phi}\bigl [
\;\;{\overline R} + 4(\nabla\phi)^2-\lambda\bigr ]
\eqno(1.2)
$$
where the overbars are there to stress the presence of a differently scaled
metric. We can remove the kinetic term for the dilaton field $\phi$ using the
field redefinition ${\overline g}_{\mu\nu}= e^{2\phi}g_{\mu\nu}$,
$\eta=e^{-2\phi}$, and Eq.~(1.2) can be recast in the more appropriate form
for the gauge formulation
$$
S_{\rm SI}={1\over 4\pi}\int d^2x {\sqrt {-g}} \Bigl (\eta R - \lambda\Bigr )
\quad . \eqno (1.3)
$$
The Lagrange multiplier enforces now the scalar curvature to vanish and the
cosmological constant $\lambda$ only appears in the equations of motion for
$\eta$. Both models (1.1) and (1.3) are known to be topological and gauge
models based on two-dimensional de Sitter$^7$ and extended Poincar\'e groups
respectively.$^9$

In this paper, we investigate $N=1$ supersymmetric extensions of the
Jackiw-Teitelboim model and the string inspired dilaton gravity model. Two
approaches are possible. The first one uses a superfield formalism
developed by Howe$^{14}$ for two dimensional superspaces and was carried over
by Chamseddine$^{15}$ in the JT
model and by Park and Strominger$^{16}$ in the SI model. The second approach
focuses on topological theories in conventional two-dimensional spacetime,
where a vanishing field strength reproduces the standard torsion and
supercurvature of supergravity. In this approach, gauge transformations
replace diffeomorphisms, Lorentz transformations and supersymmetries.
If the topological nature of the supersymmetric extension
of the JT model is known,$^{17}$ only recently Rivelles has worked out a
topological theory leading to a supersymmetric extension of the second
model.$^{18}$

The structure of the paper goes as follows:
In section II, we review the work of Chamseddine, Park and Strominger. We write
their $N=1$ supersymmetric extension in component fields.
In section III, we show how the Chamseddine action is also a topological and
gauge invariant model based on the graded de Sitter algebra OSP(1,1$\vert$1).
Actually, it is not possible to write the Park and Strominger action as a
topological theory of the $\int\langle\eta , F\rangle $-type
(without fields redefinitions). Park and Strominger's model
leads to a non-vanishing curvature and represents a challenge to be written
as a topological and gauge invariant model of an other type. Nevertheless,
in section IV, we are
able to construct other $N=1$ supersymmetric extensions described by
topological and gauge invariant models using two examples of graded
extensions of the extended Poincar\'e algebra. Rivelles's proposal is
one of them.

\bigskip
\noindent {\bf II. Supersymmetric Extension of the Jackiw-Teitelboim
and String Inspired Models. }
\medskip

Chamseddine proposed a $N=1$ supersymmetric extension of the Jackiw-Teitelboim
model in a superfield formalism
$$\ick {
S_{\rm SJT} &=-{i\over 8\pi}
\int d^2x\;d^2\theta E\Phi \Bigl ( S - \lambda' \Bigr
)\quad.
&(2.1) \cr }
$$
Our conventions are in the appendix.
If we integrate out the $\theta$ variable and eliminate the auxiliary fields
by their classical equations of motion
$$
A=\lambda' \qquad F= {1\over 2}\lambda'\phi \eqno(2.2)
$$
we recast the action in component fields
$$\ick {
S'_{\rm SJT}={1\over 4\pi}
\int d^2x \;\sqrt{-g} & \left\{  \phi \left( R + {1\over 2}
{\lambda'}^2
 - {i\over4}\lambda' \tepsilon^{\mu\nu} \chi_\mu{}^{\alpha}
 \bigl( \gamma^5\bigr )_\alpha{}^{\beta} \chi_{\nu,\beta}\right) \right.\cr
& \left. - 2i \Lambda^\alpha \left( \tepsilon^{\mu\nu}
 \bigl( \gamma^5\bigr )_\alpha{}^{\beta} D_\mu \chi_{\nu,\beta}
 + {1\over4} \lambda' \bigl( \gamma^\mu\bigr )_\alpha{}^{\beta}
 \chi_{\mu,\beta} \right)
\right\} & (2.3)\cr}
$$
where the prime on $S_{\rm SJT}$ means that we have suppressed the auxiliary
fields. Setting the fermion fields to zero reproduces $S_{\rm JT}$
with a negative cosmological constant $\lambda=-(\lambda')^2/2$.

The supersymmetric extension of the Jackiw-Teitelboim model is rather trivial
to obtain since
one substitutes for the fields the corresponding superfields. However, this is
not the case for the string inspired dilaton gravity model.
The same strategy would lead one to move the
parenthesis to the left of $\Phi$ to construct a $N=1$
supersymmetric extension of the model (1.3), that is,
$$
S=-{i\over 8\pi}
\int d^2x \;d^2\theta\; E\Bigl (\Phi S - \lambda'\Bigr )\eqno (2.4)
$$
Integrating out the $\theta$-variable and eliminating the auxiliary fields
gives the action
$$\ick {
S' & = {1\over 4\pi}
\int d^2x\; \sqrt{-g} \Bigl \{ \phi R
- {i\lambda'\over 4}\tepsilon^{\mu\nu}\chi_\mu^{\;\;\alpha}
\bigl( \gamma^5\bigr )_\alpha^{\;\;\;\beta}\chi_{\nu,\beta}
-2i \Lambda^\alpha\tepsilon^{\mu\nu}
\bigl( \gamma^5\bigr )_\alpha^{\;\;\;\beta}D_\mu\chi_{\nu,\beta}
\Bigr \} &(2.5)\cr}
$$
But, setting the fermion field to zero does not reproduce the model (1.3) since
there is no cosmological constant present.

Turning back to the original form (1.2) of the string inspired model and
replacing the fields by superfields leads indeed to a suitable $N=1$
supersymmetric extension. This was carried over by Park and Strominger
whose goal was to provide a positive energy theorem. Their
actions in superspace and in components are
$$\ick {
{\overline S}_{\rm SSI}&=-{i\over 8\pi}
\int d^2x\;d^2\theta \;E\; e^{-2\Phi}\Bigl [
S+2iD_\alpha\Phi D^\alpha\Phi - \lambda'\Bigl ]\cr
{\overline S}'_{\rm SSI} = & {1\over 4\pi} \int d^2x \;\sqrt{-g} \;e^{-2\phi}
 \Bigl \{ R + 4g^{\mu\nu}\partial_\mu\phi\partial_\nu\phi
 + {1\over4} (\lambda')^2  \cr
& + 4i \tepsilon^{\mu\nu}
 \Lambda^\alpha \bigl( \gamma^5\bigr )_\alpha^{\;\;\;\beta}
 D_\mu\chi_{\nu,\beta}-4i\Lambda^\alpha
 \bigl( \gamma^\mu\bigr )_\alpha^{\;\;\;\beta}\partial_\mu\Lambda_\beta
 - i\lambda'\Lambda^\alpha\Lambda_\alpha
 + {i\over2} \lambda'\Lambda^\alpha
 \bigl( \gamma^\mu\bigr )_\alpha{}^\beta \chi_{\mu,\beta} \cr
& -2i\Lambda^\alpha\bigl( \gamma^\nu\bigr )_\alpha^{\;\;\;\beta}
 \bigl( \gamma^\mu\bigr )_\beta^{\;\;\;\gamma}\chi_{\mu,\gamma}
 \partial_\nu\phi
 - {i\over 4}\tepsilon^{\mu\nu} (\lambda' - 2i \Lambda^\gamma \Lambda_\gamma )
 \chi_\mu{}^\alpha
 \bigl( \gamma^5\bigr )_\alpha{}^\beta \chi_{\nu,\beta}
 \Bigr\} &(2.6) \cr }
$$
where the fields should be read with overbars as in Eq.~(1.2). Setting the
fermion fields to zero reproduces ${\overline S}_{\rm SI}$ of Eq.~(1.2) with
negative cosmological constant $\lambda=-(\lambda')^2/4$.

Going from ${\overline S}_{\rm SI}$ in Eq.~(1.2) to $S_{\rm SI}$ in
Eq.~(1.3) was achieved by a Weyl transformation ${\overline g}_{\mu\nu} =
e^{2\phi} g_{\mu\nu}$. A supersymmetric
extension of $S_{\rm SI}$ can therefore be obtained by performing a
super-Weyl transformation on the supersymmetric extension of ${\overline
S}_{\rm SI}$. These super-transformations are discussed by Howe
and they act on the super-Zweibein by
$$
{\overline e}_\mu^a = e^\phi e_\mu^a \qquad
{\overline \chi}_\mu{}^\alpha = e^{{1\over 2}\phi} \bigl [
 \chi_\mu{}^\alpha + \bigl( \gamma_\mu\bigr )^{\alpha\beta}
\Lambda_\beta \bigr ] \eqno(2.7)
$$
and on the scalar field $\Phi$, they do not affect $\phi$ but its
superpartner $\Lambda_\alpha$
$$ {\overline \Lambda}_\alpha =
 e^{-{1\over 2}\phi} \Lambda_\alpha \eqno(2.8)
$$
[Note that the auxiliary fields get also transformed: $\overline{A} =
e^{-\phi} (A + 2 F)$ and $\overline{F} = e^{-\phi} F$.]
The action (2.6) then becomes
$$\ick {
S'_{\rm SSI} = & {1\over 4\pi} \int d^2x \sqrt{-g}
 \Bigl \{\; e^{-2\phi} R + {1\over4} (\lambda')^2
 + 6i e^{-2\phi} \tepsilon^{\mu\nu} \Lambda^\alpha
 \bigl( \gamma^5\bigr )_\alpha{}^\beta D_\mu \chi_{\nu,\beta} \cr
& -8i e^{-2\phi}\Lambda^\alpha\bigl(\gamma^\mu\bigr)_\alpha{}^\beta
  D_\mu\Lambda_\beta
- i \Lambda^\alpha \bigl( \gamma^\mu \gamma^\nu\bigr )_\alpha{}^\beta
 \chi_{\mu,\beta} \partial_\nu e^{-2\phi}
  - {1\over2} e^{-2\phi} \Lambda^\alpha \Lambda_\alpha g^{\mu\nu}
 \chi_\mu{}^\beta \chi_{\nu,\beta} \cr
& + 2ie^{-2\phi}\tepsilon^{\mu\nu}D_\mu\Lambda^\alpha
\bigl(\gamma^5\bigr)_\alpha{}^\beta\chi_{\nu,\beta}
-{i\over4} e^{-2\phi} \tepsilon^{\mu\nu} (\lambda' e^\phi - 4i
 \Lambda^\gamma \Lambda_\gamma) \chi_\mu{}^\alpha
 \bigl( \gamma^5\bigr )_\alpha{}^\beta \chi_{\nu,\beta}
\Bigr\}  & (2.9) \cr }
$$
\bof{$$\ick {
S'_{\rm SSI}
=&{1\over 4\pi}
\int d^2x \;e \Bigl \{\;e^{-2\phi} R + 4ie^{-2\phi}\Lambda^\alpha
\tepsilon^{\mu\nu}\bigl( \gamma^5\bigr )_\alpha^{\;\;\;\beta}
D_\mu\chi_{\nu,\beta} + {1\over4} (\lambda')^2 \cr
 & + i\lambda'\Lambda^\alpha
\bigl( \gamma^\mu\bigr )_\alpha^{\;\;\;\beta}\chi_{\mu,\beta}
+i\Lambda^\alpha\bigl( \gamma^\nu\bigr )_\alpha^{\;\;\;\beta}
\bigl( \gamma^\mu\bigr )_\beta^{\;\;\;\gamma}\chi_{\mu,\gamma}
\partial_\nu e^{-2\phi}  \cr
&+{1\over 2}e^{-2\phi}\Lambda^\alpha\Lambda_\alpha\chi_\mu^{\;\;\beta}
g^{\mu\nu}\chi_{\nu,\beta}
- 2 i \lambda' e^{-\phi}\Lambda^\alpha\Lambda_\alpha
- {i\over4} \lambda'e^{-\phi}\tepsilon^{\mu\nu}\chi_\mu^{\;\;\alpha}
\bigl( \gamma^5\bigr )_\alpha^{\;\;\;\beta}\chi_{\nu,\beta}
\Bigr\}  & (2.9) \cr }
$$}
This action now obviously reduces to Eq.~(1.3) if the fermions fields are
set to zero.

\bigskip
\noindent {\bf III. Gauge Theoretic Formulation of Supergravity Theories. }
\medskip
The lineal gravities (1.1) and (1.3) share the remarkable property to
possess a topological and gauge invariant formulation.$^9$ The main
concern of
this paper is to see whether this remains true for supersymmetric
extensions.
For the Jackiw-Teitelboim model, this formulation is based on the
two-dimensional de Sitter group. It turns out
that the same is true for the Chamseddine  extension (2.1)
provided one works with the graded de Sitter algebra OSP(1,1$\vert$1)
$$\ick {
[P_a,P_b] & = - {1\over4} (\lambda')^2 \epsilon_{ab} J  \quad
[P_a,J] = \epsilon_a{}^{b} P_b \quad
[Q_\alpha ,J] = {1\over 2} \bigl( \gamma^5\bigr )_\alpha{}^{\beta} Q_\beta
\cr
[P_a,Q_\alpha] & = {1\over4} \lambda' \bigl( \gamma_a\bigr
 )_\alpha{}^{\beta}Q_\beta \quad
\{Q_\alpha,Q_\beta\} = -2i\bigl( \gamma^a\bigr )_{\alpha\beta}P_a
+ i \lambda' \bigl( \gamma^5\bigr )_{\alpha\beta} J  & (3.1)
\cr}
$$
In that case, the (graded, see appendix) invariant non-degenerate inner product
$$
\langle P_a , P_b \rangle \equiv h_{ab} \quad
\langle J , J \rangle \equiv 4/(\lambda')^2 \quad
\langle Q_\alpha , Q_\beta \rangle \equiv - (8i/\lambda')
 \epsilon_{\alpha\beta} \eqno(3.2)
$$
is used to write the action
$$
S_{\rm GJT} = {1\over 4\pi}
\int d^2x \, \epsilon^{\mu\nu} \langle \eta , F_{\mu\nu} \rangle
\eqno(3.3)
$$
where $F = dA + A^2$ is the strength field associated with the gauge field
$$
A_\mu = e^a_\mu P_a - \omega_\mu J + {1\over 2} \, \chi_\mu{}^{\alpha}
\bigl( \gamma^5\bigr )_\alpha{}^{\beta} Q_\beta \eqno (3.4)
$$
and $\eta = \eta^a P_a + \eta^J J + \eta^\alpha Q_\alpha$ is a world scalar
with value in the graded algebra~(3.1).
This action is explicitly topological and gauge invariant. Remark that it
is defined on an ordinary two-dimensional surface.
In components, it writes
$$\ick{
S_{\rm GJT} = {1\over 4\pi}\int d^2x\,  \Bigl \{
& \eta_a \epsilon^{\mu\nu} \left( \partial_\mu e_\nu^a - \epsilon^a{}_b
\omega_\mu e_\nu^b - {i\over4} \chi_\mu{}^\alpha \bigl(
\gamma^a \bigr)_\alpha{}^\beta \chi_{\nu,\beta} \right) \cr
& - {4\over(\lambda')^2} \eta^J \epsilon^{\mu\nu} \left( \partial_\mu
 \omega_\nu + {1\over 8} (\lambda')^2 \epsilon_{ab} e^a_\mu e^b_\nu
 + {i\over8} \lambda' \chi_\mu{}^{\alpha} \bigl( \gamma^5\bigr
 )_\alpha{}^\beta \chi_{\nu,\beta} \right) \cr
& + {4i\over\lambda'} \eta_\alpha \left( \epsilon^{\mu\nu} D_\mu
 \chi_\nu{}^\beta \bigl( \gamma^5\bigr )_\beta{}^{\alpha}
 + {1\over 4} e\;\lambda' \chi_\mu{}^\beta \bigl( \gamma^\mu\bigr
 )_\beta{}^{\alpha} \right)
\Bigr \} \quad. & (3.5) \cr }
$$
If we solve the constraint enforced by $\eta_a$, we get a relation between
$\omega_\mu$ and $e_\mu^a,\, \chi_\mu{}^\alpha$
$$
\omega_\mu = - e_{\mu,a} \tepsilon^{\rho\sigma} \partial_\rho e_\sigma^a
+{i\over 2}\chi_\mu^{\;\;\alpha}\bigl( \gamma^5 \gamma^\nu\bigr
)_\alpha{}^\beta \chi_{\nu,\beta} \eqno(3.6)
$$
identical to the one between the spin-connection, the Zweibein and the
gravitino [see Eq.~(A.13)] obtained from the standard kinematic constraints
on the supertorsion.

If we use the relation (3.6) and we set $\eta^J = (\lambda')^2\phi/2$ and
$\eta^\alpha = - \lambda' \Lambda^\alpha/2$ in the action (3.5), it reduces to
the action of Chamseddine  (2.3).
In this gauge formulation the supergravity transformations
on the fields are replaced by some gauge transformations of the
gauge fields. Indeed, following
Howe, but replacing the auxiliary fields as in Eq. (2.2), the supergravity
transformations [see Eqs (A.14)] are
$$
\delta e_\mu^{\;\;a} =i\tau^\alpha\bigl (\gamma^a\bigr )_\alpha^{\;\;\beta}
\chi_{\mu,\beta} \qquad
\delta \omega_\mu  = - {i\over 2}\lambda '
\tau^\alpha\bigl (\gamma^5\bigr )_\alpha{}^\beta
\chi_{\mu,\beta} \qquad
\delta \chi_\mu^{\;\;\alpha} =2D_\mu\tau^\alpha + {\lambda '\over2}
\bigl(\gamma_\mu\bigr)^{\alpha\beta}\tau_\beta \eqno(3.7)
$$
for the Zweibein, the spin-connection and the gravitino
fields respectively. In the gauge theory, the
transformations on the gauge fields are obtained with
$$
A' =  U^{-1}A U + U^{-1}d U \eqno (3.8)
$$
and the infinitesimal transformation
$U=1+\tau^\alpha\bigl(\gamma^5\bigr )_\alpha^{\;\;\beta}Q_\beta$ reproduces
local supersymmetries of Eq.~(3.7).

In the SI case, the supersymmetric extension proposed by Park and
Strominger -- after the super-Weyl transformation~(2.7) -- possesses
quartic interaction terms and thus does not fit with the form~(3.3).
Moreover, variation of the action $S'_{\rm SSI}$ of Eq. (2.9)
with respect to $e^{-2\phi}$ and
$\Lambda_\alpha$
gives two equations: One containing both the curvature $R$ and the
gravitino kinetic term $D_\mu\chi_\nu$ and another one that gives
an expression for $D_\mu\chi_\nu$. Substituting the
equation for $D_\mu\chi_\nu$ into the first produces
$$\ick {
R= &\;\;\; -16i\Lambda^\alpha\bigl(\gamma^\mu\bigr )_\alpha{}^\beta
   D_\mu\Lambda_\beta
   -{1\over 2}\Lambda^\alpha\Lambda_\alpha\bigl [ \chi_\mu{}^\beta
   g^{\mu\nu}\chi_{\nu,\beta} +2\tepsilon^{\mu\nu}\chi_\mu{}^\beta
   \bigl(\gamma^5\bigr)_\beta{}^\gamma\chi_{\nu,\gamma}\bigr ]\cr
& +2i \Lambda^\alpha\bigl ( \gamma^\mu\gamma^\nu\bigr )_\alpha{}^\beta
  \chi_{\mu,\beta} \partial_\nu\phi
-i\partial_\nu \Bigl [
  \Lambda^\alpha\bigl ( \gamma^\mu\gamma^\nu\bigr )_\alpha{}^\beta
  \chi_{\mu,\beta} \Bigr ] \cr
& -2i {e^{2\phi}\over {\sqrt {-g}} }D_\mu \Bigl [
  \Lambda^\alpha{\sqrt{-g}}\tepsilon^{\mu\nu}e^{-2\phi}\bigl (
  \gamma^5\bigr )_\alpha{}^\beta\chi_{\nu,\beta}\Bigr ]
  - {i\over 8} \lambda' e^{\phi} \tepsilon^{\mu\nu}
   \chi_\mu^{\;\;\;\alpha}
   \bigl ( \gamma^5\bigr )_\alpha^{\;\;\;\beta}\chi_{\nu,\beta}
& (3.9) \cr }
$$
Setting the fermion fields to zero gives $R=0$, which was needed to construct a
topological and gauge invariant model. However, if the fermions are present,
there is no way (without introducing new fields) to absorb the RHS of
Eq.~(3.9) in a redefinition of the spin-connection in order to get a vanishing
curvature condition.

We then turn to the strategy of building supersymmetric extensions
by considering topological theories of the $\int
\langle\eta, F\rangle$-type. We will present in the next section
examples of topological and gauge invariant models that are based on the
simplest graded extension of the extended Poincar\'e algebra and we will
see that they actually differ from the action~(2.9).
\bigskip
\noindent {\bf IV.A Graded extension of the extended Poincar\'e algebra}
\medskip

Motivated by the contraction of the graded de Sitter algebra (3.1), Rivelles
obtained a graded extension of the extended Poincar\'e algebra. However, the
algebra he presents is not the result of the contraction, but rather, a
modification of it that satisfies the Jacobi identity
$$\ick {
[P_a,P_b] & = \epsilon_{ab}I \quad [P_a, J] = \epsilon_a^{\;\;b}P_b \quad
  [P_a,Q_\alpha] = {1\over 2} \bigl( \gamma_a\bigr )_\alpha{}^\beta U_\beta
  \cr
[Q_\alpha ,J] & ={1\over 2}\bigl( \gamma^5\bigr )_\alpha^{\;\;\;\beta}Q_\beta
  \quad [U_\alpha,J]
  ={1\over 2}\bigl( \gamma^5\bigr )_\alpha^{\;\;\;\beta}U_\beta \quad
  [Q_\alpha,K] = {1\over 2}
  \bigl( \gamma^5\bigr )_\alpha^{\;\;\;\beta}U_\beta & (4.\hbox{A}1) \cr
\{Q_\alpha,Q_\beta\} & = - 2i \bigl( \gamma^a\bigr )_{\alpha\beta}P_a
 + 2i \bigl( \gamma^5\bigr )_{\alpha\beta}K \quad
 \{U_\alpha , Q_\beta\} = -2i \bigl( \gamma^5\bigr )_{\alpha\beta}I
 \cr}
$$
It possesses the (graded) invariant and non-degenerate inner product
$$
\langle P_a , P_b \rangle = h_{ab} \quad \langle J , I \rangle = -1 \quad
\langle K , K \rangle = 1  \quad
\langle Q_\alpha , U_\beta \rangle = -4i \epsilon_{\alpha\beta}
\eqno (4.\hbox{A}2)
$$
[see conventions in Eq.~(A.1)].

A gauge theory based on the algebra (4.\hbox{A}1) is built from the
strength field
$F = dA + A^2$ associated to the gauge field
$$
A_\mu = e^a_\mu P_a - \omega_\mu J + a_\mu I + b_\mu K
+ {1\over2} \chi_\mu{}^{\alpha} \bigl( \gamma^5\bigr )_\alpha{}^{\beta}
Q_\beta + \xi_\mu{}^{\alpha} U_\alpha \eqno (4.\hbox{A}3)
$$
and a Lagrange multiplier $\eta = \eta^a P_a + \eta(J) J + \eta(I) I+
\eta(K) K +
\eta^\alpha(Q) Q_\alpha + \eta^\alpha(U) U_\alpha$, which transforms
(like $F = dA + A^2$) under the adjoint representation. The action
$$
S_{\rm GEP} = {1\over 4\pi}
\int d^2x \;\epsilon^{\mu\nu} \langle \eta , F_{\mu\nu} \rangle
\eqno(4.\hbox{A}4)
$$
is obviously topological and invariant under
gauge transformations generated by the
algebra (4.\hbox{A}1).
Local supersymmetries are obtained by gauge transformations in the
additional directions $K$, $Q_\alpha$ and $U_\alpha$. The infinitesimal
transformation $U = 1 + \zeta K + \tau^\alpha
\bigl(\gamma^5\bigr)_\alpha{}^\beta Q_\beta + \sigma^\alpha U_\alpha$ generates
on $A_\mu$ the transformations
$$\ick {
\delta e^a_\mu &= i\tau^\alpha\bigl
(\gamma^a\bigr )_\alpha{}^\beta\chi_{\mu,\beta} \quad
\delta\omega_\mu =0 \quad
\delta\chi_\mu{}^\alpha =2D_\mu\tau^\alpha \cr
\delta a_\mu &= 2i\tau^\alpha\xi_{\mu,\alpha} -i \sigma^\alpha \chi_\alpha
 \quad
\delta b_\mu = \partial_\mu \zeta + i \tau^\alpha\bigl
(\gamma^5\bigr )_\alpha{}^\beta \chi_{\mu,\beta} & (4.\hbox{A}5) \cr
\delta\xi_\mu{}^\alpha &= D_\mu \sigma^\alpha + {1\over2}
\;\tau^\beta \bigl(\gamma^5\gamma_\mu\bigr)_\beta{}^\alpha - {1\over2}
\;\tau^\alpha b_\mu + {1\over4} \zeta \chi_\mu{}^\alpha \cr }
$$
The first three are Howe's supergravity transformations [see Eq. (A.14)]
with the auxiliary field $A=0$ taking a different value than
in Park and Strominger's approach.
The $\eta$-field also transforms in the usual way, $\eta'=U^{-1}\eta U$
$$
\ick {
\delta\eta^a&=2i\tau^\alpha\bigl (\gamma^a\gamma^5\bigr )_\alpha{}^\beta
\eta_\beta (Q) \quad
\delta\eta (J) =0 \quad
\delta\eta^\alpha (Q) = - {1\over2} \tau^\alpha\eta (J)  \cr
\delta\eta (I)&=2i\tau^\alpha\eta_\alpha (U) + 2 i \sigma^\alpha
 \bigl(\gamma^5\bigr)_\alpha{}^\beta \eta_\beta(Q) \quad
\delta\eta (K) =-2i\tau^\alpha\eta_\alpha (Q) &(4.\hbox{A}6) \cr
\delta\eta^\alpha (U) &= {1\over 2}\tau^\beta\bigl (\gamma^5\gamma^a
\bigr )_\beta{}^\alpha\eta_a - {1\over 2}\tau^\alpha\eta (K) + {1\over2}
\zeta \eta^\beta(Q) \bigl(\gamma^5\bigr)_\beta{}^\alpha - {1\over2}
\sigma^\beta \bigl(\gamma^5\bigr)_\beta{}^\alpha \eta(J)
\cr}
$$

The equations of
motion are $F = 0$, or in components,
$$\ick {
F^a(P) &=
 d e^a - \epsilon^a{}_{b}\omega e^b
 -{i\over4} \chi^\alpha
 \bigl( \gamma^a\bigr )_\alpha{}^{\beta}\chi_\beta = 0 \cr
F(J) &= -d \omega = 0 \cr
F(I) &=
 d a +{1\over 2}\epsilon_{ab}e^a e^b
 -i \chi^\alpha \xi_\alpha =0 \cr
F(K) &=
 d b - {i\over4}
 \chi^\alpha \bigl( \gamma^5\bigr )_\alpha{}^{\beta}
 \chi_\beta =0 & (4.\hbox{A}7) \cr
F^\alpha(Q) &= \bigl(
 d\chi^\beta - {1\over 2}\omega
 \bigl( \gamma^5\bigr )^{\beta\gamma} \chi_\gamma
 \bigr ) {1\over2} \bigl( \gamma^5\bigr )_{\beta}{}^\alpha =0 \cr
F^\alpha(U) &=
 d \xi^\alpha - {1\over 2}\omega
 \bigl( \gamma^5\bigr )^{\alpha\beta} \xi_\beta
 + {1\over4} e^a
 \epsilon_{ab} \bigl( \gamma^b\bigr )^{\alpha\beta} \chi_\beta
 - {1\over4} b \chi^\alpha = 0
\cr}
$$
and $(D\eta) \equiv d\eta + [ A , \eta ] = 0$, or in components,
$$\ick {
(D \eta)_a(P) = & d \eta_a - \epsilon_{ab} \omega \eta^b
 - \epsilon_{ab} e^b \eta(J) + i \epsilon_{ab} \chi^\alpha
 \bigl( \gamma^b\bigr )_\alpha{}^\beta \eta_\beta(Q)
 =0 \cr
(D \eta)(J) = & d \eta(J) =0 \cr
(D \eta)(I) = & d \eta(I) + e^a \epsilon_a{}^b \eta_b
 - 2i \xi^\alpha \bigl( \gamma^5\bigr )_\alpha{}^\beta \eta_\beta(Q)
 - i \chi^\alpha \eta_\alpha(U) =0 \cr
(D \eta)(K) = & d \eta(K) + i \chi^\alpha \eta_\alpha(Q)
 =0 & (4.\hbox{A}8) \cr
(D \eta)_\alpha(Q) = & d \eta_\alpha(Q) - {1\over 2} \omega
 \bigl( \gamma^5\bigr )_\alpha{}^\beta \eta_\beta(Q)
 + {1\over4} \chi_\alpha \eta(J) =0 \cr
(D \eta)_\alpha(U) = & d \eta_\alpha(U) - {1\over 2} \omega
 \bigl( \gamma^5\bigr )_\alpha{}^\beta \eta_\beta(U)
 + {1\over2} \left( b \bigl( \gamma^5 \bigr)_\alpha{}^\beta
 - e^a \bigl( \gamma_a \bigr )_\alpha{}^\beta \right) \eta_\beta(Q)
\cr
 & + {1\over4} \bigl( \gamma^a \bigr )_\alpha{}^\beta \chi_\beta
 \epsilon_{ab} \eta^b
 + {1\over 4} \chi_\alpha \eta(K)
 - {1\over2} \bigl( \gamma^5\bigr )_\alpha{}^\beta \xi_\beta \; \eta(J)
 =0 \cr}
$$
%Setting the fermion fields to zero, we recover the model~(1.3), written in
%a gauge formulation,$^9$
%giving thus an alternative supersymmetric extension to the SI model.

Quite remarkably, it is possible to eliminate some of the fields by
finding relations between them, which solve part of the
differential equations. We get
in the Lagrange multiplier
$$\ick{\displaystyle
\eta^a = & \tepsilon^{\mu\nu} \partial_\mu \eta(I) e_\nu^a
 + i \tepsilon^{\mu\nu}
e^a_\mu\chi_\nu{}^\alpha\eta_\alpha (U) \cr
\eta(J) = & - \lambda & (4.\hbox{A}9) \cr
\eta(K) = & -{2i\over\lambda} \eta^\alpha(Q) \eta_\alpha(Q) - 4 \lambda'
\cr}
$$
and in the gauge field
$$\ick{
\omega_\mu = & - e_{\mu,a} \tepsilon^{\rho\sigma} \partial_\rho e^a_\sigma
 + {i\over2} \chi_\mu{}^\alpha \bigl( \gamma^5 \gamma^\nu\bigr
 )_\alpha{}^\beta \chi_{\nu,\beta}  \cr
b_\mu = & - {i\over\lambda} \chi_\mu{}^\alpha \bigl( \gamma^5\bigr
 )_\alpha{}^\beta \eta_\beta(Q) \cr
\chi_\mu{}^\alpha = & {4\over\lambda} D_\mu \eta^\alpha(Q) & (4.\hbox{A}10) \cr
\xi_\mu{}^\alpha = & - {i\over2\lambda^2} \eta^\gamma(Q) \eta_\gamma(Q)
 \bigl(\gamma^5\bigr)^{\alpha\beta}  \chi_{\mu,\beta} + {1\over\lambda}
 \bigl( \gamma_\mu\gamma^5\bigr )^{\alpha\beta} \eta_\beta(Q)
  \cr}
$$
where $\lambda$ and $\lambda'$ are constants of integration. The fields
$b_\mu$ and $\xi_\mu{}^\alpha$ are also determined up to quantities,
which can be gauged away by a transformation in the directions $K$ and
$U_\alpha$.
As a consequence, the resulting transformations of the fields in
(4.A9) and (4.A10) are consistent with the ones given in (4.A5) and (4.A6)
if we choose $\zeta = -(2i/\lambda) \tau^\alpha
\bigl(\gamma^5\bigr)_\alpha{}^\beta \eta_\beta(Q)$ and $\sigma^\alpha =
(i/\lambda^2) \eta^\gamma(Q) \eta_\gamma(Q) \tau^\beta
\bigl(\gamma^5\bigr)_\beta{}^\alpha$
$$\ick {
\delta e^a_\mu &= i\tau^\alpha\bigl
(\gamma^a\bigr )_\alpha{}^\beta\chi_{\mu,\beta} \quad
\delta\eta (I) =2i\tau^\alpha\eta_\alpha (U)
\quad \delta\eta^\alpha\bigl (Q\bigr )={\lambda\over 2}\tau^\alpha
& (4.\hbox{A}11) \cr
\delta\eta^\alpha (U) &= - {1\over 2}\tau^\beta\bigl(\gamma^\mu
 \bigr)_\beta{}^\alpha \biggl( \partial_\mu \eta(I) - i \chi_\mu{}^\beta
\eta_\beta(U) \biggr) + 2 \lambda' \tau^\alpha \left( 1 +
 {i\over\lambda\lambda'} \eta^\beta(Q) \eta_\beta(Q) \right)
\cr}
$$

\bof{If we use these relations, the Lagrangian reduces to
$$\ick{
 S'_{\rm GEP} = {1\over4\pi} \int d^2x & \sqrt{-g} \biggl\{
  \left( -{1\over2} \eta(I) \right) R - \lambda \cr
  &+ i \tepsilon^{\mu\nu}
  \bigl( -\lambda \chi_\mu{}^\alpha \xi_{\nu,\alpha} + 4 \eta_\alpha(U)
  F_{\mu\nu}^\alpha(Q) + 4 \eta_\alpha(Q) F_{\mu\nu}^\alpha(U) \bigr)
  \biggr\}
 & (4.\hbox{A}11) \cr }
$$
Remark that even if $\xi_\mu{}^\alpha$ is given explicitly in (4.A9),
we cannot eliminate it from the Lagrangian since we do not know how to
solve for $\eta_\alpha(Q)$. Variation of action~(4.A11) or}
Substitution of relations (4.\hbox{A}9) and (4.\hbox{A}10) in the
equations of motion (4.\hbox{A}7) and  (4.\hbox{A}8) provides a new set of
differential equations;
they represent the dynamics of the gauge theory (4.\hbox{A}4)
where the remaining unconstrained fields of the theory are $e_\mu^a,
\eta(I), \eta^\alpha(Q), \eta^\alpha(U)$
$$\ick {
& R(e_\mu^a, \chi_\nu{}^\alpha ) =0 \cr
& ( \nabla_\mu \partial_\nu - g_{\mu\nu} \nabla_\rho \partial^\rho ) \eta(I)
+ \lambda g_{\mu\nu} =
i ( \delta_\mu^\rho \delta_\nu^\sigma + \delta_\mu^\sigma \delta_\nu^\rho
 - 2 g_{\mu\nu} g^{\rho\sigma} ) \chi_\rho{}^\alpha
 \bigl(\gamma_\sigma\bigr)_\alpha{}^\beta \eta_\beta(Q) \cr
& \hskip 15mm + i \chi_\mu{}^\alpha \chi_{\nu,\alpha} \left( \lambda' +
{i\over\lambda} \eta^\beta(Q) \eta_\beta(Q) \right)
- {i\over2} g_{\mu\nu} {1\over\sqrt{-g}} D_\rho ( \sqrt{-g} g^{\rho\sigma}
\chi_\sigma)^\alpha \eta_\alpha(U) \cr
& \chi_\mu{}^\alpha = {4\over\lambda} D_\mu \eta^\alpha(Q) & (4.\hbox{A}12) \cr
& D_\mu\eta^\alpha (U) - \chi_\mu{}^\alpha \left( \lambda' + {i\over \lambda}
 \eta^\beta (Q) \eta_\beta (Q) \right) - \bigl(\gamma_\mu\bigr)^{\alpha\beta}
\eta_\beta (Q)
+ {1\over 4}\chi_\mu{}^\beta\bigl ( \gamma^\nu\bigr )_\beta{}^\alpha
\partial_\nu \eta (I) \cr
& \hskip 3in - {i\over 4}\chi_\mu{}^\beta \bigl (
\gamma^\nu\bigr )_\beta{}^\alpha\chi_\nu{}^\gamma\eta_\gamma (U) =0
\cr }
$$
[The covariant derivative $\nabla_\mu$ contains the non-symmetric Christoffel
symbol
defined by $e^a_\rho \Gamma^\rho_{\mu\nu} \equiv \partial_\mu e_\nu^a -
\epsilon^a{}_b \omega_\mu e_\nu^b - (i/4) \chi_\mu{}^\alpha
\bigl(\gamma^a\bigr)_\alpha{}^\beta \chi_{\nu,\beta}$.]
The gauge field
$a_\mu$ is not considered
here. Its equation of motion (4.\hbox{A}7) can always be locally integrated
because $da$ is equated to a two-form, hence closed in two dimensions.

The corresponding bosonic SI theory possesses classical solutions
parametrized by the cosmological constant $\lambda$ and a ``mass'' $M$.$^9$
The local symmetries we have
added will be useful if they turn out to be symmetries of the above
configurations. We have
already broken the ones associated with $K$ and $U_\alpha$ in the
reduction~(4.A9), (4.A10). Let us consider the following family of
solutions
$$\ick {
& \eta^\alpha\bigl(Q\bigr)={\rm constant} \qquad (\chi_\mu{}^\alpha=0) \cr
&\partial_\mu\eta^\alpha\bigl (U\bigr ) =
\bigl(\gamma_\mu\bigr )^{\alpha\beta}\eta_\beta\bigl (Q\bigr ) =
{\rm constant}. \cr
&g_{\mu\nu}=h_{\mu\nu}\qquad
( \partial_\mu \partial_\nu - h_{\mu\nu} \partial_\rho \partial^\rho ) \eta(I)
 + \lambda h_{\mu\nu} = 0 &(4.{\hbox {A}}13) \cr}
$$
which reproduces the usual bosonic configurations for $g_{\mu\nu}$ and
$\eta(I)$ described in Ref. 9, and reduces the action (4.A4) to the
form given in Eq.~(1.3).
It is invariant under the transformations
$(4.{\hbox {A}}11)$.
This means that the solutions (4.A13) possess the Howe's supersymmetry
(4.A11), which allows us to construct a conserved charge like in Park and
Strominger's work.
\bof {A configuration with vanishing fermion fields
will be
symmetric under the remaining $Q$-symmetry if the following equations are
satisfied with $\tau \neq 0$
$$
\delta\eta^\alpha (Q)| = {1\over2} \lambda\tau^\alpha=0 \quad
\delta\eta^\alpha (U)| = {1\over2} \bigl(\gamma^\mu\bigr)^{\alpha\beta}
\tau_\beta \partial_\mu\eta (I) + 2 \lambda' \tau^\alpha = 0 \quad
\delta\chi_\mu{}^\alpha| =2D_\mu\tau^\alpha=0 \eqno(4.\hbox{A}13)
$$
where the ``$|$'' stress the evaluation at zero Fermi fields. The first
equality implies $\lambda=0$. The model we are
considering has vanishing curvature hence the metric is flat and from the
last equality we deduce that $\tau^\alpha$ is a constant spinor
with two independent components. Finally, the second equality tells us that
$\lambda'=0$ and $\eta(I) = {\rm constant}$.
Clearly, $\lambda\neq 0$ breaks the
supersymmetry. We conclude that either the bosonic vacuum is trivial
($\lambda = 0$)
and supersymmetry is preserved, or, the vacuum is not trivial (the
interesting case) and supersymmetry is broken by the vacuum.
}
\bigskip
\noindent {\bf IV.B Minimal Graded extension of the extended
Poincar\'e algebra}
\medskip
We now turn to another proposal for topological and supersymmetric extension
of the SI model, which looks simpler and also shows a set of
bosonic solutions with a non trivial symmetry.
There is a minimal graded algebra consistent with the Jacobi identities
and containing the bosonic generators of the extended Poincar\'e algebra (3.1).
Introducing the fermionic generator $Q_\alpha$ as the only newly added charge,
we define the following algebra
$$\ick {
[P_a,P_b]&=\epsilon_{ab} \; I\quad [P_a,J]=\epsilon_a^{\;\;b}P_b \cr
\{Q_\alpha,Q_\beta\}&=-2i [(1-\gamma_5)\gamma^a]_{\alpha\beta}
P_a +i
\bigl( \gamma^5\bigr )_{\alpha\beta}\;I &(4.\hbox{B}1)\cr
[Q_\alpha, J]&= {1\over 2}\bigl( \gamma^5\bigr )_\alpha^{\;\;\;\beta}
Q_\beta \quad
[Q_\alpha,P_a]= [(1-\gamma_5)\gamma_a]_\alpha{}^\beta
Q_\beta \cr }
$$
with inner product given by
$$
\langle P_a,P_b\rangle =h_{ab}\quad \langle J,I\rangle =-1\quad
\langle Q_\alpha , Q_\beta\rangle =2i
\epsilon_{\alpha\beta} \eqno(4.\hbox{B}2)
$$

A gauge theory based on the algebra (4.\hbox{B}1) will be simpler then the
theory based on the previous algebra (4.\hbox{A}1) because there are
less fields needed to
write a topological and gauge invariant action. The gauge field is
$$
A_\mu=e^a_\mu P_a - \omega_\mu J + a_\mu I + {1\over 2}
\chi_\mu{}^\alpha\bigl (\gamma^5\bigr )_\alpha{}^\beta Q_\beta
 \eqno(4.\hbox{B}3)
$$
and the Lagrange multiplier $\eta = \eta^a P_a + \eta(J) J + \eta(I) I+
\eta^\alpha Q_\alpha$. The action is again
$$\ick {
S_{\rm MGEP}&={1\over 4\pi}\int d^2x \epsilon^{\mu\nu}\langle
\eta,F_{\mu\nu}\rangle
&(4.\hbox{B}4) \cr }
$$
where $F=dA+A^2$, and it is topological and invariant
under gauge transformations. In particular, in the $Q$-direction with $U =
1 + \tau^\alpha \bigl(\gamma^5\bigr)_\alpha{}^\beta Q_\beta$
$$\ick {
\delta e^a_\mu &= i\tau^\alpha [(1-\gamma_5)\gamma^a]_\alpha{}^\beta
\chi_{\mu,\beta} \quad
\delta\omega_\mu =0 \quad
\delta a_\mu = {i\over2} \tau^\alpha\bigl (\gamma^5\bigr )_\alpha{}^\beta
\chi_{\mu,\beta} \cr
\delta\chi_\mu{}^\alpha&=2D_\mu\tau^\alpha
+ 2 \tau^\beta [(1-\gamma_5)\gamma_\mu]_\beta{}^\alpha &
(4.\hbox{B}5) \cr }
$$
and for the $\eta$-fields
$$\ick{
\delta \eta^a&=-2i\tau^\alpha [(1-\gamma^5)\gamma^a]_\alpha{}^\beta
\eta_\beta \quad
\delta \eta(J) = 0 \quad
\delta\eta (I) =-i\tau^\alpha\eta_\alpha \cr
\delta\eta^\alpha&= -{1\over 2}\tau^\alpha\eta (J) + \tau^\beta
[(1-\gamma^5)\gamma^a]_\beta{}^\alpha \eta_a  &(4.\hbox{B}6) \cr }
$$

The equations of motion of the model (4.B4) are $F=0$, or in components,
$$\ick {
F^a(P) &= d e^a - \epsilon^a{}_b \omega e^b
 - {i\over 4} \chi^\alpha [(1-\gamma_5)\gamma^a]_\alpha{}^\beta \chi_\beta =0
 \cr
F(J) &= - d \omega = 0 \cr
F(I) &= d a + {1\over 2} \epsilon_{ab} e^a e^b - {i\over 8}
 \chi^\alpha \bigl(\gamma^5\bigr)_\alpha{}^\beta \chi_\beta = 0
 & (4.\hbox{B}7) \cr
F^\alpha (Q) &= {1\over 2} \left( D \chi^\beta
 - e^a [(1-\gamma_5)\gamma_a]^{\beta\gamma} \chi_\gamma \right)
 \bigl(\gamma^5\bigr)_\beta{}^\alpha = 0
 \cr}
$$
and $(D\eta) \equiv d\eta +[A,\eta]=0$, or in components,
$$\ick {
(D \eta)_a(P) &=
 d \eta_a - \epsilon_a{}^b \omega \eta_b -\epsilon_{ab}
 e^b \eta(J) + i \chi^\alpha
 [(1-\gamma_5)\gamma^a]_\alpha{}^\beta
 \eta_\beta =0 \cr
(D \eta)(J)&= d \eta(J)=0 \cr
(D \eta)(I)&= d \eta(I) + e^a \epsilon_a{}^b \eta_b
 + {i\over 2} \chi^\alpha \eta_\alpha = 0 &(4.\hbox{B}8) \cr
(D \eta)_\alpha(Q) &= D\eta_\alpha + {1\over 4} \chi_\alpha
 \eta(J) + e^a [(1-\gamma_5)\gamma_a]_\alpha{}^\beta \eta_\beta
 + {1\over 2} \eta^a [(1-\gamma_5)\gamma_a]_\alpha{}^\beta \chi_\beta
 = 0 \cr }
$$
We remark the systematic appearance of combinations we will denote
$\psi_\alpha^{\rm R,L} = {1\over2} (1\pm \gamma^5)_\alpha{}^\beta \psi_\beta$.
\bof{
We recognize in the transformation of the Zweibein a chiral supersymmetry.
The transformations (4.B5) coincide with those of Howe if the auxiliary field
is taken to be $A = -4$ and $\tau_\alpha^{\rm L} = 0$ ($\tau_- = 0$)
[see Eq. (A.14)].
}
As in the previous example, it is possible to eliminate some of the
fields by replacing part of these differential equations by relations between
the fields. In the gauge fields, we can impose
$$\ick {
\omega_\mu &= - e_{\mu,a} \tepsilon^{\rho\sigma} \partial_\rho e_\sigma^a
 + {i\over 2} \chi_\mu^{{\rm R},\alpha} \bigl( \gamma^5\gamma^\nu\bigr
 )_\alpha{}^\beta \chi_{\nu,\beta}^{\rm R} &(4.\hbox{B}9) \cr }
$$
and in the Lagrange multiplier
$$\ick {
\eta^a &= \tepsilon^{\mu\nu} e_\nu^a \left( \partial_\mu\eta(I) + {i\over 2}
 \chi_\mu{}^\alpha \eta_\alpha \right) \cr
\eta(J)& = - \lambda = {\rm constant } & (4.\hbox{B}10) \cr }
$$
\bof{
Again, substitution of the above relations in the
Lagrangian reduces it to
$$\ick{
 S'_{\rm MGEP} = {1\over4\pi} \int d^2x \sqrt{-g} \biggl\{ &
  \left( - {\eta(I)\over2} \right) R - \lambda \cr
 & + i \tepsilon^{\mu\nu} ( {\lambda\over8} \chi_\mu{}^\alpha \bigl( \gamma^5
  \bigr)_\alpha{}^\beta \chi_{\nu,\beta} - 2 \eta_\alpha
  F_{\mu\nu}^\alpha(Q) ) \biggr\} & (4.\hbox{B}11) \cr  }
$$
}
The equations of motion
provide the dynamics for the remaining unconstrained fields
$e_\mu^a, \chi_\mu{}^\alpha, \eta (I), \eta^\alpha$ of the gauge theory
(4.\hbox{B}4) upon substituting the relations (4.B9),(4.B10)
$$\ick {
& R \left( e^a_\mu , \chi_\mu{}^{{\rm R,}\alpha}
 \bigl(\gamma^\nu\bigr)_\alpha{}^\beta \chi_{\nu,\beta}^{\rm R} \right) =0 \cr
& D \chi^{{\rm L,}\alpha} - 2 e^a \bigl(\gamma_a\bigr)^{\alpha\beta}
\chi^{\rm R}{}_\beta =0 \cr
& D \chi^{{\rm R,}\alpha} = 0 \cr
& ( \nabla_\mu \partial_\nu - g_{\mu\nu} \nabla_\rho \partial^\rho ) \eta(I)
+ \lambda g_{\mu\nu} =
{3i\over2} ( \delta_\mu^\rho \delta_\nu^\sigma + \delta_\mu^\sigma
  \delta_\nu^\rho - 2 g_{\mu\nu} g^{\rho\sigma} ) \chi_\rho^{{\rm R,}\alpha}
  \bigl(\gamma_\sigma\bigr)_\alpha{}^\beta \eta_\beta^{\rm R}
  & (4.\hbox{B}11) \cr
& \hskip 1cm + {i\over4} g_{\mu\nu} {1\over\sqrt{-g}} D_\rho ( \sqrt{-g}
  g^{\rho\sigma} \chi_\sigma^{\rm R})^\alpha \eta_\alpha^{\rm L}
+ {i\over4} g_{\mu\nu} {1\over\sqrt{-g}} D_\rho ( \sqrt{-g} g^{\rho\sigma}
  \chi_\sigma^{\rm L})^\alpha \eta_\alpha^{\rm R} \cr
& \hskip 1cm + {i\over4} \chi_\mu^{{\rm R,}\alpha} \chi_{\nu,\alpha}^{\rm L}
+ {i\over4} \chi_\mu^{{\rm L,}\alpha} \chi_{\nu,\alpha}^{\rm R} \cr
& D \eta_\alpha^{\rm L} - {\lambda\over4} \chi^{\rm L}{}_\alpha +
 2 e^a \bigl(\gamma_a\bigr)_\alpha{}^\beta  \eta_\beta^{\rm R}
 - \chi^{{\rm R,}\beta} \bigl(\gamma^\nu\bigr)_{\beta\alpha}
 \bigl[ \partial_\nu \eta(I) + {i\over 2} \chi_\nu{}^{{\rm R,}\gamma}
 \eta_\gamma^{\rm L} + {i\over 2} \chi_\nu{}^{{\rm L,}\gamma}
 \eta_\gamma^{\rm R} \bigr]
 =0 \cr
& D \eta_\alpha^{\rm R} - {\lambda\over4} \chi^{\rm R}{}_\alpha = 0
\cr }
$$
We do not consider the $a_\mu$ field for the same reason as in the previous
model. These equations are symmetric under
$$\ick {
\delta e^a_\mu &= 2i\tau^{{\rm R,}\alpha} \bigl(\gamma^a\bigr)_\alpha{}^\beta
\chi^{\rm R}_{\mu,\beta} \cr
\delta\chi_\mu^{{\rm R,}\alpha}&=2D_\mu\tau^{{\rm R,}\alpha} \quad
\delta\chi_\mu^{{\rm L,}\alpha} =2D_\mu\tau^{{\rm L,}\alpha}
+ 4 \tau^{{\rm R,}\beta} \bigl(\gamma_\mu\bigr)_\beta{}^\alpha \cr
\delta\eta (I) &= - i \tau^{{\rm R,}\alpha} \eta^{\rm L}_\alpha
 - i \tau^{{\rm L,}\alpha} \eta^{\rm R}_\alpha & (4.\hbox{B}12) \cr
\delta\eta^{{\rm R,}\alpha} &= {\lambda\over 2} \tau^{{\rm R,}\alpha} \quad
\delta\eta^{{\rm L,}\alpha} = {\lambda\over 2} \tau^{{\rm L,}\alpha}
 + 2 \tau^{{\rm R,}\beta} \bigl(\gamma^a\bigr)_\beta{}^\alpha \eta_a
 \cr }
$$
$\eta_a$ being given in Eq.~(4.B10). We recover Howe's supergravity
transformations for the right chirality sector [see Eq.~(A.14) with $A=0$],
but the left chirality sector follows another type of transformations.

The bosonic solutions described in Ref.~9 are still among the solutions
$$\ick{
\chi^{{\rm R,}\alpha} &= 0 \qquad
\eta_\alpha^{\rm L} = {\rm constant} \qquad
\chi_\mu^{{\rm L,}\alpha} = {8\over\lambda}
 \bigl(\gamma_\mu\bigr)_\alpha{}^\beta \eta_\beta^{\rm R} = {\rm constant}
 \cr
g_{\mu\nu} &= h_{\mu\nu} \qquad
( \partial_\mu \partial_\nu - h_{\mu\nu} \partial_\rho \partial^\rho ) \eta(I)
 + \lambda h_{\mu\nu} = 0
 & (4.\hbox{B}13) \cr}
$$
and
one can verify that the action (4.B4) reduces to the one discussed in Ref.~9
when the gravitino field $\chi_\mu{}^\alpha$ is evaluated as above.
This subset of solutions is invariant under (4.B12) with $\tau^{\rm
R}_\alpha = 0$ and $\tau^{\rm L}_\alpha = {\rm constant}$. The condition
$\tau^{\rm R}_\alpha=0$ is necessary to insure consistency of the third
equation in $(4.\hbox{B}13)$. This means that the right chirality
transformations identified as Howe supergravity transformations are
broken by the solutions (4.B13). Nevertheless, the
remaining symmetries is still one of (4.B13) and allows us to construct
the corresponding conserved charge.
\bigskip
\noindent {\bf Conclusions.}
\medskip

Four models for two dimensional supergravity theories have been discussed
here. The $N=1$ supersymmetric extension of JT-model given by Chamseddine,
and three models that are $N=1$ supersymmetric extension of the string
inspired dilaton gravity model. We emphasized that the first model is
topological and a gauge invariant
$\int \langle\eta, F\rangle$-theory based on the graded de Sitter algebra
OSP(1,1$\vert$1), the local supersymmetries being reproduced by some gauge
transformations. We showed that the
$\int \langle\eta, F\rangle$ action is indeed the superfield
action written by Chamseddine in Howe's superfield formalism for supergravity
theories. The second model we
analyzed is the Park and Strominger's proposal. They write a $N=1$
supersymmetric
extension following the lines of Howe. We arrived at the conclusion
that it is not a $\int \langle\eta,F\rangle$-theory and has
non-vanishing curvature. Although,
it does not mean that it is not an other type of topological theory, we believe
that it represents a challenge to write it as such.
Then, we turn to the strategy of building two supersymmetric extensions
of string inspired dilaton gravity models from topological and gauge invariant
$\int \langle\eta,F\rangle$ actions based on graded extension of the
extended Poincar\'e algebras. The first of the two topological
models was presented
first by Rivelles. The algebra he uses to define the GEP-model
was inspired by a contraction of the graded de Sitter algebra. However, it is
not a contraction but rather an algebra that looks like it and closes under
the Jacobi identity.
The second model is the minimal
graded extended Poincar\'e model. There, only one charge $Q$ is added.
The connection to supergravity theories is
made by comparing local supersymmetries given by
Howe's formalism for two-dimensional supergravity theories
with some gauge transformations of the proposed topological models. Although,
the first model present full supersymmetry, the connection for the second
model is made possible only for half the supersymmetry; we have a chiral
supersymmetry. We conclude that they are then supergravity theories that
extend the string inspired dilaton gravity because upon setting the
gravitino
field to zero, we recover the bosonic model. In the first model,
we find field configurations with vanishing gravitino that
remain supersymmetric. The second model offers a similar result, however,
in this case,
the configurations with vanishing right gravitino break
half the gauge symmetry. In both models, a conserved charge can be constructed
for the bosonic solutions of Ref.~9.
Its consequence in the physics of the system deserve
another study.
\bof{either the vacuum is trivial (the cosmological constant vanishes) and the
supersymmetry is preserved, or, the vacuum is not trivial
and the supersymmetry is completely broken. The second model offers a more
interesting
result. The bosonic solutions break only half of the gauge symmetry
and the remaining symmetry leads to a conserved charge.
}
\bigskip
\noindent {\bf Appendix }
\medskip
We use the following convention for the inner product presented in section III
and IV
$$
\langle \xi^A Q_A , \eta^B Q_B \rangle = (-)^{\deg(Q_A)\deg(Q_B)} \xi^A
\eta^B \langle Q_A, Q_B \rangle \eqno (\hbox{A}.1)
$$
[$Q_A$ indicates any generator of the algebra and
$\deg(Q_A) = 0$ if $Q_A$ is an even generator and $\deg(Q_A) = 1$ if $Q_A$
is an odd generator].

The rest of the appendix deals with various other
conventions. The indices are denoted in the paper like this:
$a,b$ are tangent space indices, $\mu, \nu$ are spacetime indices and $\alpha,
\beta$ are spinor indices. The bosonic metric is
$$h_{ab}={\rm diag}(-1,1) \quad
\epsilon_{ab}=-\epsilon_{ba} \quad \epsilon_{01}=1\eqno (\hbox{A}.2)$$
and the contraction of two anti-symmetric tensors is given by $$
\epsilon_{ab}\epsilon^{bc}=\delta_a^{\;\;c} \eqno (\hbox{A}.3)
$$
The fermionic metric is
$$\epsilon_{\alpha\beta}=\epsilon^{\alpha\beta} \quad
\epsilon_{12}=1=-\epsilon_{21} \quad \epsilon_{11}=\epsilon_{22}=0
\eqno (\hbox{A}.4)$$
and contractions are given by
$$
\epsilon_{\alpha\beta}\epsilon^{\beta\gamma}=-\delta_\alpha{}^\gamma \quad
\epsilon_{\alpha\beta}\epsilon^{\alpha\beta}=2 \eqno (\hbox{A}.5)
$$
The spin indices are raised and lowered by the fermionic metric
$$\psi^\alpha=\epsilon^{\alpha\beta}\psi_\beta \quad
\psi_\alpha=\psi^\beta \epsilon_{\beta\alpha}\eqno (\hbox{A}.6)$$
and the product of two Grassmann variables reduces to
$$
\theta^\alpha\theta^\beta=-{1\over 2}\epsilon^{\alpha\beta}\theta^\gamma
\theta_\gamma   \eqno (\hbox{A}.7)
$$
We choose our $\gamma$-matrices real and such that they satisfy
$$
\bigl( \gamma^a\bigr )_\alpha^{\;\;\;\beta}
  \bigl( \gamma^b\bigr )_\beta^{\;\;\;\gamma}
=\eta^{ab}\delta_\alpha^{\;\;\;\gamma} - \epsilon^{ab}
  \bigl( \gamma^5\bigr )_\alpha^{\;\;\;\gamma}\eqno (\hbox{A}.8) $$
With $\gamma^5= \gamma^0\gamma^1$, we deduce the useful relations
$$
[\gamma^a, \gamma^b ]=-2\epsilon^{ab}
\gamma^5 \quad
[\gamma^a, \gamma^5 ]=2\epsilon^{ab}
\gamma_b \quad
\gamma^a\gamma^5 = \epsilon^{ab} \gamma_b
\eqno (\hbox{A}.9)
$$
An explicit representation is
$$
\bigl( \gamma^0\bigr )_\alpha{}^\beta =
\pmatrix{0&1\cr -1&0\cr} \quad
\bigl( \gamma^1\bigr )_\alpha{}^\beta =
\pmatrix{0&1\cr 1&0\cr} \quad
\bigl( \gamma^5\bigr )_\alpha{}^\beta =
\pmatrix{1&0\cr 0&-1\cr} \eqno (\hbox{A}.10)
$$

The superfield formalism is given in the work of Howe.$^{14}$ Here are
the needed formulae. The scalar superfield is defined by
$$
\Phi=\phi+i\theta^\alpha\Lambda_\alpha +{i\over 2}\theta^\alpha\theta_\alpha
F \eqno (\hbox{A}.11)
$$
where the $\phi$-field is a scalar field, $\Lambda_\alpha$ the fermion field
associated to the scalar field, and $F$ an auxiliary field to be
determined by the model under investigation.
The supersymmetric covariant derivative,
$D_\alpha=\partial_\alpha+i\theta^\beta
\bigl( \gamma^a\bigr )_{\beta\alpha}\partial_a$,
is used in Eq.~(2.6).
The superfields associated to the
graviton multiplet are the supervolume element $E$, and the supercurvature
$S$ given by
$$\ick {
E&=e\bigl [ 1+{i\over 2}\theta^\alpha
\bigl( \gamma^\mu\bigr)_\alpha^{\;\;\;\beta}
\chi_{\mu,\beta}+
\theta^\alpha\theta_\alpha\bigl ({i\over 4}A+{1\over
8}{\tilde\epsilon}^{\mu\nu}
\chi_\mu^\alpha\bigl( \gamma^5\bigr )_\alpha^{\;\;\;\beta}\chi_{\nu,\beta}
\bigr)\bigr] \cr
S&=A+\theta^\alpha\Psi_\alpha+{i\over 2}\theta^\alpha\theta_\alpha C
&(\hbox{A}.12)\cr
&C=-R-{1\over 2}\chi_\mu^{\;\;\;\alpha}
\bigl( \gamma^\mu\bigr )_\alpha^{\;\;\;\beta}
\Psi_\beta +{i\over 4}{\tilde\epsilon}^{\mu\nu}\chi_\mu^{\;\;\;\alpha}
\bigl( \gamma^5\bigr )_\alpha^{\;\;\;\beta}\chi_{\nu,\beta}A-{1\over 2}A^2
\cr
&\Psi_\alpha= -2i{\tilde\epsilon}^{\mu\nu}
\bigl( \gamma^5\bigr )_\alpha^{\;\;\;\beta}
D_\mu\chi_{\nu,\beta} -{i\over 2}\bigl( \gamma^\mu\bigr )_\alpha^{\;\;\;\beta}
\chi_{\mu,\beta}A \cr}
$$
where the superfields $E$ and $S$ are expressed in terms of the Zweibein
$e_\mu^a$, the gravitino $\chi_\mu{}^\alpha$ and an auxiliary field $A$.
{}From these fields the spacetime covariant derivative, the spin-connection
and the curvature are constructed following Howe's work. We list these
relations here
$$\ick {
g_{\mu\nu}&=e^a_\mu e^b_\nu h_{ab} \quad
 e=-{1\over 2} \epsilon^{\mu\nu}e_\mu^ae_\nu^b
 \epsilon_{ab} = {\sqrt {-g}} \cr
\gamma_\mu&=e_\mu{}^a\gamma_a \quad
 \gamma^\mu = E^\mu_a \gamma^a = -{1\over e}\epsilon^{\mu\nu}\epsilon_{ab}
 e^b_\nu \gamma^a \cr
\omega_\mu&=-e_{\mu,a}{\tilde\epsilon}^{\nu\rho}\partial_\nu e_\rho^a
 + {i\over 2}\chi_\mu^{\;\;\;\alpha}
 \bigl( \gamma^5 \gamma^\nu\bigr )_\alpha{}^\beta \chi_{\nu,\beta}
&(\hbox{A}.13) \cr
R&=-2 {1\over e} \epsilon^{\mu\nu}\partial_\mu\omega_\nu \cr
D_\mu\chi_{\nu,\alpha}&=
 \partial_\mu\chi_{\nu,\alpha}-{1\over 2}\omega_\mu
 \bigl( \gamma^5\bigr )_\alpha^{\;\;\;\beta}\chi_{\nu,\beta}
\cr}
$$
Note the sign convention for the scalar curvature $R$.
Howe deduces also the supergravity transformations after having imposed
some kinematics constraints on the supertorsion
$$\ick{
\delta e_\mu{}^a & =i \tau^\alpha\bigl (\gamma^a\bigr )_\alpha{}^\beta
\chi_{\mu,\beta} \quad
\delta \omega_\mu = -{i\over2} A \tau^\alpha
\bigl(\gamma^5\bigr)_\alpha{}^\beta \chi_{\mu,\beta} \quad
\delta\chi_\mu{}^\alpha=2D_\mu\tau^\alpha - {1\over 2} \tau^\beta
 \bigl (\gamma_\mu\bigr )_\beta{}^\alpha A \cr
\delta \phi & = 0 \quad \hskip 25mm
\delta \Lambda_\alpha = \tau^\beta  \bigl(\gamma^\mu\bigr)_{\beta\alpha}
\left( \partial_\mu \phi + {i\over2} \chi_\mu{}^\gamma \Lambda_\gamma
\right) + \tau_\alpha F &
(\hbox{A}.14) \cr}
$$
Of some interest are also the antisymmetric tensors
$$
{\tilde \epsilon}^{\mu\nu} ={1\over e}\epsilon^{\mu\nu} = \epsilon^{ab}
E_a^\mu E_b^\nu \quad
{\tilde \epsilon}_{\mu\nu} =e\epsilon_{\mu\nu} = \epsilon_{ab} e_\mu^a
e_\nu^b \quad
\tepsilon^{\mu\nu} \tepsilon_{\nu\rho} = \delta^\mu_\rho
\eqno(\hbox{A}.15)
$$
and the action of a Weyl transformation on the scalar curvature
$$
g_{\mu\nu}=e^{2\phi}g'_{\mu\nu} \quad\longrightarrow\quad
{\sqrt {-g}}R={\sqrt {-g'}}R'-2 \partial_\mu\left(\sqrt {-g'}
g^{\prime\mu\nu} \partial_\nu \phi \right)
\eqno(\hbox{A}.16)
$$

\vfill\eject

\medskip
\noindent {\bf References }
\medskip
\medskip
\item {1.} C. Teitelboim, Phys. Lett. {\bf 126B} (1983) 41;
{\it Quantum Theory of Gravity}, S.~Christensen, ed. (Adam Hilger, Bristol,
1984); R. Jackiw, {\it Quantum Theory of Gravity}, S. Christensen, ed.
(Adam Hilger, Bristol, 1984); Nucl. Phys. {\bf B252} (1985) 343.
\medskip
\item {2.} E. Witten, Comm. Math. Phys. {\bf 118} (1988) 411.
\medskip
\item {3.} J. Labastida, M. Pernici, and E. Witten, Nucl. Phys. {\bf B310}
(1988) 611.
\medskip
\item {4.} H. Verlinde, {\it The Sixth Marcel Grossman Meeting on General
Relativity}, H. Sato, ed. (World Scientific Singapore, 1992); C. Callan,
S. Giddings, A. Harvey, and A.~Strominger, Phys. Rev. {\bf D45}
(1992) 1005.
\medskip
\item {5.} R.B. Mann and T.G. Steele, Class. Quan. Grav. {\bf 9} (1992) 475.
\medskip
\item {6.} A.E. Sikkema and R.B. Mann, Class. Quan. Grav. {\bf 8} (1991) 219.
\medskip
\item {7.} T. Fukuyama and K. Kamimura, Phys. Lett. {\bf 160B} (1985) 259;
 K. Isler and C.~Trugenberger, Phys. Rev. Lett. {\bf 63} (1989) 834;
 A. Chamseddine and D. Wyler, Phys. Lett. {\bf B228} (1989) 75.
\medskip
\item {8.} D. Montano and J. Sonnenschein, Nucl. Phys. {\bf B313} (1989) 258.
\medskip
\item {9.} D. Cangemi and R. Jackiw, Phys. Rev. Lett. {\bf 69} (1992) 233.
\medskip
\item {10.} A. Ach\`ucarro and P.K. Townsend, Phys. Lett. {\bf B180}
(1986) 89; E. Witten, Nucl. Phys. {\bf B311} (1988) 46.
\medskip
\item {11.} M. Ba\~nados, C. Teitelboim, and J. Zanelli, Phys. Rev. Lett.
{\bf 69} (1992) 1849;
 M. Ba\~nados, M. Henneaux, C. Teitelboim, and J. Zanelli,
 IAS preprint (Nov. 1992) submitted to Phys. Rev. D.
\medskip
\item {12.} D. Cangemi, M. Leblanc, and R.B. Mann, MIT-CTP\#2162 preprint
(Nov. 1992) submitted to Phys. Rev. D.
\medskip
\item {13.} R.~Brooks, Nucl. Phys. {\bf B320} (1989) 440.
\medskip
\item {14.} P.S. Howe, J. Phys. A: Math. Gen. {\bf 12} (1979) 393.
\medskip
\item {15.} A. H. Chamseddine, Phys. Lett. {\bf B258} (1991) 97.
\medskip
\item {16.} Y. Park and A. Strominger, UCSBTH-92-39 preprint, hep-th/9210017
(1992).
\medskip
\item {17.} D. Montano, K. Aoaki, and J. Sonnenschein, Phys. Lett. {\bf B247}
(1990) 64; E.~D'Hoker, Phys. Lett. {\bf B264} (1991) 101.
\medskip
\item {18.} V.O. Rivelles, IFUSP-P/1025 preprint (Dec. 1992).
\medskip

\end